\documentstyle[aps,prl,epsf,floats,rotate]{revtex}
\newcommand{\etal}{\it et al.\rm}
\newcommand{\rt}{\rightarrow}
\def\Journal#1#2#3#4{{#1} {\bf #2}, #3 (#4)}

\def\NIMA{{\em Nucl. Instrum. Methods} A}
\def\NPB{{\em Nucl. Phys.} B}
\def\PLB{{\em Phys. Lett.}  B}
\def\PRL{\em Phys. Rev. Lett.}
\def\PRD{{\em Phys. Rev.} D}
\def\ZPC{{\em Z. Phys.} C}

\begin{document}

%\preprint{\vbox{\hbox{BIHEP-EP1-xx-0?\hfill}
%                \hbox{UH511-xxx-xx\hfill}}}
%@twocolumnfalse\endcsname 
%\twocolumn

% Titlepage and abstract are here

\title{\boldmath MEASUREMENT OF THE TOTAL CROSS SECTION FOR HADRONIC 
PRODUCTION BY $e^+e^-$ ANNIHILATION AT ENERGIES BETWEEN 2.6-5 GEV
\cite{support}}

% Following can be inserted into revtex document.
\author{
J.~Z.~Bai,$^1$   Y.~Ban,$^4$      J.~G.~Bian,$^1$
G.~P.~Chen,$^1$  H.~F.~Chen,$^{9}$  
J.~Chen,$^2$ 
J.~C.~Chen,$^1$  Y.~Chen,$^1$ Y.~B.~Chen,$^1$  Y.~Q.~Chen,$^1$
% ...
\cite{atNU0}
Y.~Y.~Wang,$^1$  
C.~L.~Wei,$^1$   N.~Wu,$^1$       Y.~G.~Wu,$^1$
D.~M.~Xi,$^1$    X.~M.~Xia,$^1$   P.~P.~Xie,$^1$    Y.~Xie,$^1$
Y.~H.~Xie,$^1$   G.~F.~Xu,$^1$    S.~T.~Xue,$^1$
J.~Yan,$^1$      W.~G.~Yan,$^1$   C.~M.~Yang,$^1$   C.~Y.~Yang,$^1$
H.~X.~Yang,$^1$  J.~Yang,$^1$     
W.~Yang,$^2$
X.~F.~Yang,$^1$  M.~H.~Ye,$^1$    S.~W.~Ye,$^{9}$
Y.~X.~Ye,$^{9}$   C.~S.~Yu,$^1$    C.~X.~Yu,$^1$     G.~W.~Yu,$^1$
Y.~H.~Yu,$^3$    Z.~Q.~Yu,$^1$    C.~Z.~Yuan,$^1$   Y.~Yuan,$^1$
B.~Y.~Zhang,$^1$ C.~Zhang,$^1$    C.~C.~Zhang,$^1$ D.~H.~Zhang,$^1$ 
Dehong~Zhang,$^1$
H.~L.~Zhang,$^1$ J.~Zhang,$^1$    J.~W.~Zhang,$^1$  L.~Zhang,$^1$
L.~S.~Zhang,$^1$ P.~Zhang,$^1$
Q.~J.~Zhang,$^1$ S.~Q.~Zhang,$^1$ X.~Y.~Zhang,$^5$  Y.~Y.~Zhang,$^1$
D.~X.~Zhao,$^1$  H.~W.~Zhao,$^1$  Jiawei~Zhao,$^{9}$ J.~W.~Zhao,$^1$
M.~Zhao,$^1$     W.~R.~Zhao,$^1$  Z.~G.~Zhao,$^1$   J.~P.~Zheng,$^1$
L.~S.~Zheng,$^1$ Z.~P.~Zheng,$^1$ B.~Q.~Zhou,$^1$   G.~P.~Zhou,$^1$
H.~S.~Zhou,$^1$  L.~Zhou,$^1$     K.~J.~Zhu,$^1$    Q.~M.~Zhu,$^1$
Y.~C.~Zhu,$^1$   Y.~S.~Zhu,$^1$   B.~A.~Zhuang$^1$
\\ (BES Collaboration)}

\address{
$^1$Institute of High Energy Physics, Beijing 100039, People's Republic of
 China\\
$^2$Colorado State University, Fort Collins, Colorado 80523\\
$^3$Hangzhou University, Hangzhou 310028, People's Republic of China\\
$^4$Peking University, Beijing 100871, People's Republic of China\\
$^5$Shandong University, Jinan 250100, People's Republic of China\\
$^6$Shanghai Jiaotong University, Shanghai 200030, People's Republic of China\\
$^7$Stanford Linear Accelerator Center, Stanford, California 94309\\
$^8$University of Hawaii, Honolulu, Hawaii 96822\\
$^{9}$University of Science and Technology of China, Hefei 230026,
People's Republic of China\\
$^{10}$University of Texas at Dallas, Richardson, Texas 75083-0688}

\date{\today}

\maketitle

%\onecolumn
\begin{abstract}
Using the upgraded Beijing Spectrometer (BESII), 
we have measured the total cross section for $e^+e^-$ annihilation 
into hadronic final states at
center-of-mass energies of 2.6, 3.2, 3.4, 3.55, 4.6 and 
5.0 GeV.  Values of $R$, $\sigma(e^+e^-\rightarrow
\mbox{hadrons})/\sigma(e^+e^-\rightarrow\mu^+\mu^-)$, are determined.
\end{abstract}
%\twocolumn
\nopagebreak
\vspace*{0.3cm}

% Begin main paper here
%\twocolumn
%\section{Introduction}
The lowest order cross section for
$e^+e^-\rightarrow\gamma^*\rightarrow \mbox{hadrons}$ is
usually parameterized in terms of the ratio $R$, which is defined as
$R=\sigma(e^+e^- \rightarrow \mbox{hadrons})/\sigma(e^+e^-\rightarrow
\mu^+\mu^-)$,
where the denominator is the lowest-order QED cross section,
$\sigma (e^+e^- \rightarrow \mu^+\mu^-) = \sigma^0_{\mu \mu}=
4\pi \alpha^2 / 3s$.
This ratio has been measured by many experiments over
the center-of-mass (cm) energy range from the hadron production 
threshold to the $Z$ pole \cite{pdg}. The 
measured $R$ values are, in general, consistent
with theoretical predictions, and provide an impressive confirmation
of the hypothesis of three color degrees of freedom for quarks.

However, the existing $R$ measurements for cm energies below 5 GeV
were performed 17 to 25 years ago~\cite{2,3,4,5,gamma2,MarkI,pluto}
and have average experimental 
uncertainties of about 15\%~\cite{BP}. Uncertainties 
in the values of $R$ in this energy region limit the precision of
the QED running coupling constant evaluated at the mass of the $Z$ boson,
$\alpha(M^2_{Z})$, which in turn limits the precision of the determination
of the Higgs mass from radiative corrections in the Standard
Model~\cite{BP,6,7,8,Eidelman,swartz,9}.
Measurements of $R$, particularly for cm energies below the $J/\psi$ mass,
are also required for the interpretation of the muon $(g-2)$
measurement at Brookhaven~\cite{BP,6,7,8,Eidelman,swartz,9}.
About 50\% and 20\% of the error in $\alpha(M^2_{Z})$ and
$a_{\mu}=(g-2)/2$, respectively, are due to the uncertainty of the
values of $R$ in the 2-5~GeV cm energy region~\cite{9}. 

%\section{Detector}

In this letter, we report measurements of $R$ at cm energies of
2.6, 3.2, 3.4, 3.55, 4.6 and 5.0 GeV. 
The measurements were carried out with the
BESII, which is a conventional solenoidal detector that is described
in detail in Ref.~\cite{10}.  Upgrades include the replacement of
the central drift chamber with a vertex chamber (VC) composed of 12
tracking layers organized around a beryllium beam pipe.  This chamber
provides a spatial resolution of about 90 $\mu$m.  The barrel
time-of-flight counter (BTOF) was replaced with a new array of 48
plastic scintillators that are read out by fine mesh photomultiplier
tubes situated in the 0.40~T magnetic field volume, providing 180 ps
resolution.  A new main drift chamber (MDC) has 10 superlayers, each
with four sublayers of sense wires.  It provides $dE/dx$ information for
particle identification and has a momentum resolution of $\sigma_p/p=1.8
\%\sqrt{(1+p^2)}$ for charged tracks with momentum $p$ in GeV.  The
sampling-type barrel electromagnetic calorimeter (BSC), which covers
80\% of 4$\pi$ solid angle, consists of 24 layers of self-quenching
streamer tubes interspersed with lead and with each layer having  560
tubes. The BSC has an energy
resolution of $\sigma_E/E=21 \%/\sqrt{E}$ ($E$ in GeV) and a spatial
resolution of 7.9 mrad in $\phi$ and 3.6 cm in $z$.  The outermost
component of BESII is a $\mu$ identification system consisting of three
double layers of proportional tubes interspersed in the iron flux return of
the magnet.  These measure coordinates along the muon trajectories with
resolutions of 3 cm and 5.5 cm in $r\phi$ and $z$, respectively.

%\section{Trigger}
Triggers are formed from signals derived from the BTOF, VC, MDC, and BSC, and 
referenced in time to signals from a beam pickup electrode located upstream
of the detector \cite{10}.
Event categories are classified according
to numbers of charged and neutral tracks seen at the trigger level.
For beam crossings with charged tracks, two trigger topologies are
utilized: in the first, we require at least one hit
in the 48 BTOF counter array, one track in the VC and MDC,
and at least 100 MeV of energy deposited in the BSC; in the second, we require
back-to-back hits in the BTOF counter with one
track in the VC and two tracks in the MDC.  For the neutral track trigger,
we require that
the sum of the deposited energy of the tracks in two adjacent towers of the
BSC is greater than 80 MeV in the first level trigger and that the total energy
deposited in BSC from all sources is greater than 800 MeV in the second level
trigger.  A tower in the BSC is one tube in $\phi$ (11 mrad) by 24
layers radially.

%\section{Data Analysis}

The value of $R$ is determined from the number of observed hadronic events
($N^{obs}_{had}$) by the relation
\begin{eqnarray*}
R=\frac{ N^{obs}_{had} - N_{bg} - \sum_{l}N_{ll} - N_{\gamma\gamma} }
{ \sigma^0_{\mu\mu} \cdot L \cdot \epsilon_{had} \cdot \epsilon_{trg}
\cdot (1+\delta)}, 
\end{eqnarray*}
where $N_{bg}$ is the number of beam associated background events;
$\sum_{l}N_{ll},~(l=e,\mu,\tau)$ and $N_{\gamma\gamma}$ are the
numbers of misidentified 
lepton-pairs from one-photon and two-photon processes events; $L$ is
the integrated luminosity; $\delta$ is the
radiative correction; and $\varepsilon_{had}$ and
$\epsilon_{trg}$ represent the detection and trigger efficiency for
hadronic events.

%\subsection{Run Stability and Performance}

The $J/\psi$ resonance is a convenient source of large numbers of
hadronic events.  A sample of 1.5 M 
$J/\psi$ events, accumulated intermittently throughout the
experimental running period, was used for monitoring the
detector performance.
These data indicate that the detector components and
triggers remained stable throughout the run \cite{Vancouver}.

%\subsection{Hadronic Event Selection}

The goal of hadronic event selection is to distinguish single-photon
hadron production from other processes.  The following track-level
selection criteria are used to define good charged tracks:

\begin{itemize}
\item $|\cos \theta| < 0.84$, where $\theta$ is the track polar angle;
\item The track must have a reasonable three-dimensional helix fit;
\item Distances of closest approach to the beam in the transverse plane
      and along the beam axis are less than 2.0 and 18 cm,
      respectively;
\item $p < p_{beam} + (5 \times \sigma_{p})$, where $p$ and $p_{beam}$
      are the momenta of the track and the beam, respectively, and
      $\sigma_p$ is the momentum resolution for charged tracks
      with $p=p_{beam}$;
\item $E < 0.6 E_{beam}$, where $E$ is the energy in the BSC 
      that is associated with the track, and $E_{beam}$
      is the beam energy;
\item A track must not be definitely identified as an electron or a muon;
\item $2 < t < t_{p} + (5 \times \sigma_{t})$ (in ns), where $t$ and
      $t_{p}$ are the time-of-flight for the track and a nominal
      time-of-flight calculated for the track assuming a proton
      hypothesis, respectively, and $\sigma_{t}$ is the BTOF time
      resolution.
\end{itemize}

After the track-level selection, a further event-level selection is
applied:

\begin{itemize}
\item At least two charged tracks, with at least one good track
satisfying the requirements listed above;
\item The total deposited energy in the BSC $> 0.28 E_{beam}$.
\end{itemize}

A further selection scheme is required based on the number of good
tracks in the event.  For three or more prong events, the only
additional requirement is that all the charged tracks not be positive
(to remove beam-gas events).  However, two-prong events must be
distinguished from cosmic ray and lepton pair events, requiring
in addition:

\begin{itemize}
\item The two tracks must not be back-to-back;

%\begin{eqnarray*}
%|\theta_{1} + \theta_{2} - 180^{\circ} | > 10^{\circ}, | |\phi_{1} -
% \phi_{2}| - 180^{\circ} | > 4^{\circ}; \nonumber
%\end{eqnarray*}

\item There must be at least two isolated neutral tracks that have
more than 100 MeV of energy and are at least 15$^{\circ}$ from the
closest charged track in azimuthal angle.

%\begin{eqnarray*}
%E_{\gamma} > 100 \mbox{MeV},
%|\phi_{\gamma} - \phi_{c}| > 15^{\circ}, \nonumber
%\end{eqnarray*}

%\noindent
%where $\phi_{c}$ is the $\phi$ angle of
%either of the two charged tracks.
\end{itemize}

%The performance of the event selection routines was checked by visual
%scans, which were carried out for both the selected and rejected
%hadronic events, the separated-beam data, and the Monte Carlo events.
%\noindent A typical hadronic event that passed the selection cuts is shown in
%Fig.~\ref{fig:event}.  We determined the error associated with the
%hadronic selection criteria from changes in the event yield caused by
%varying the selection requirements.
% and from the results of the visual scan.

%\begin{figure}[htb]
%\epsfysize=2.8in
%\centerline{\rotate[r]{\epsfbox{event_2.ps}}}
%\caption{A typical hadronic event display from 3.55 GeV data.}
%\label{fig:event}
%\end{figure}

%\subsection{Background Subtraction}

There are three major types of background to be considered.  One type,
consisting of cosmic rays, Bhabha, dimuon events and some two-photon
process events, is directly removed by the event selection.
The second, consisting of tau-pair production events and residual
lepton-pair events from two-photon processes, is subtracted out statistically.

The most serious sources of background in the hadronic event
sample are beam-gas and beam-wall interactions.
To understand these,
separated beam data
were taken at each energy point, and single beam data were accumulated at
3.55 GeV.  Most of the beam
associated background events are rejected by a vertex cut.  The
salient features of the beam associated background are that their
tracks are very much along the beam pipe direction, the energy
deposited in BSC is small, and most of the tracks are protons.  The
same hadronic event selection criteria are applied to the
separated-beam data, and the number of separated-beam events $N_{sep}$
surviving these criteria are obtained.  The number of the beam
associated background events $N_{bg}$ in the corresponding hadronic
event sample is given by $N_{bg}=f \times N_{sep}$, where $f$ is
the ratio of the product of the pressure at the
collision region times the integrated beam currents for colliding beam
runs and that for the separated beam runs.

%\subsection{Luminosity}

The integrated luminosity is determined using large-angle Bhabha events
with the following selection criteria, using only BSC information:

\begin{itemize}
\item Two clusters in the BSC with largest deposited energy in the
      polar angle $|\cos \theta| \le 0.55$;
\item Each cluster with energy $> 1.0$ GeV (for 3.55 GeV data, scaled
      for other energy points);
\item $2^{\circ} < | |\phi_{1} - \phi_{2}| - 180^{\circ} | < 16^{\circ}$,
      where $\phi_{1}$ and $\phi_{2}$ are the azimuthal angles of the
      clusters. The $2^{\circ}$ cut removes $e^+ e^- \rt \gamma
      \gamma$ events.

\end{itemize}
A cross check using only $dE/dx$ information from the MDC
to identify electrons was generally
consistent with the BSC measurement; the difference was taken into
account in the overall systematic error of 2.1-2.8\%.

%\subsection{Hadronic Detection Efficiency}

\begin{figure}[htb]
\epsfysize=2.4in
\centerline{\epsfbox{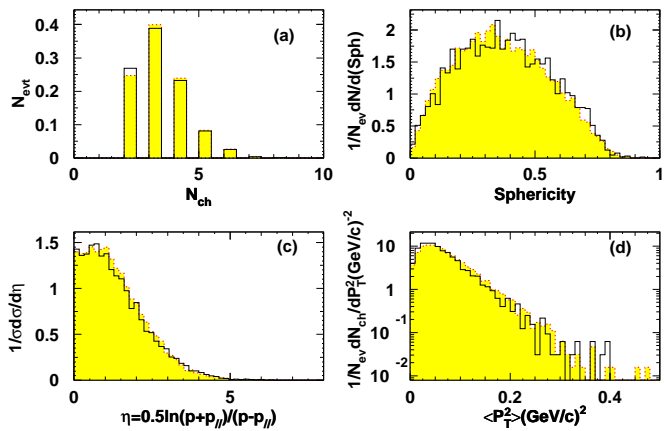}}
\caption{Comparison of hadronic event shapes between data (shaded
region) and Monte Carlo (histogram).  (a) Multiplicity; (b) Sphericity;
(c) Rapidity; (d) Transverse momentum.}
\label{fig:lund}
\end{figure}

The detection efficiency for hadronic events is determined via a Monte
Carlo simulation using the JETSET7.4 event generator\cite{11}.
Parameters in the generator are tuned \cite{qi} using a 40 k
hadronic event sample collected near 3.55 GeV for the tau mass measurement
done by the previous 
experiment at BESI\cite{12}.  The parameters of the generator are adjusted to
reproduce distributions of kinematic variables such as multiplicity,
sphericity, transverse momentum, etc.  Fig.~\ref{fig:lund} shows these
distributions for the real and simulated event samples.
The parameters have also been obtained using the 2.6 GeV data ($\approx$ 5
k events).  The difference between the two parameter sets and between
the data and the Monte Carlo data based on these parameter sets is
used to determine a systematic error of 1.9-3.2\% in the hadronic
efficiency.
 
The trigger efficiencies are measured
by comparing the
responses to different trigger requirements
in special runs taken at the $J/\psi$ resonance.
From the trigger 
measurements, the efficiencies
for Bhabha, dimuon and hadronic events are
determined to be 99.96\%, 99.33\% and 99.76\%,
respectively. As a cross check, the trigger information
from the 2.6 and 3.55 GeV data samples
are used to provide independent measurements of
the trigger efficiencies.
These are consistent with the efficiencies determined from
the $J/\psi$ data.
The errors in the
trigger efficiencies for Bhabha and hadronic events
are less than $\pm$0.5\%.

%\subsection{Initial State Radiation}

Radiative corrections determined using four different
schemes \cite{13,14,15,24} agreed with each other to within 1\%
below charm threshold.  Above charm threshold, where resonances are
important, the agreement is within $1\sim3\%$.  The major uncertainties
common to all models are due to errors in previously measured
$R$-values and in the choice of values
for the resonance parameters. For the measurements
reported here, we
use the formalism of Ref.~\cite{15} and include the differences with
the other schemes in the systematic error of 2.2-4.1\%.

%\subsection{Results}

\begin{figure}[!htb]
\epsfysize=2.7in
\centerline{\epsfbox{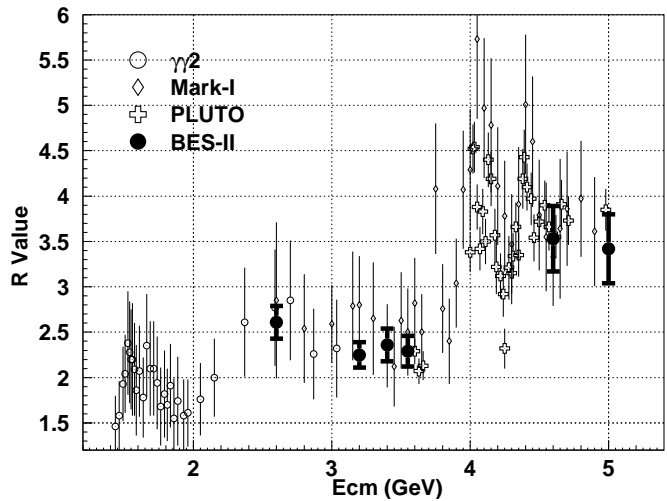}}
\vskip -.1 cm
\caption{Plot of $R$ values vs $E_{cm}$.}
\label{fig:rvalue}
\end{figure}

The $R$ values obtained at the six energy points are shown in
Table~\ref{tab:rvalue} and graphically displayed in
Fig.~\ref{fig:rvalue}.  A breakdown of contributions to the
systematic errors is given in Table~\ref{tab:rsyst}. The largest 
systematic error is due to the hadronic event selection
and is determined to be 3.8-6.0\% by varying the selection criteria.
The systematic
errors on the measurements below 4.0 GeV are similar and are a measure
of the amount of error common to all points.  We have also
done the analysis including only events with greater than two
charged tracks; although the statistics are smaller, the results
obtained agree well with the
results shown here. 
The $R$ values for $E_{cm}$ below 4 GeV are in good agreement with
results from
$\gamma \gamma 2$ \cite{gamma2} and Pluto \cite{pluto} but are below
those from Mark I \cite{MarkI}.
Above 4 GeV,
our values are consistent with previous measurements.
%Our combined statistical and systematic errors are much smaller than
%previous experiments in this energy region.

%\vspace{-0.05in}
\begin{table}[!h]
\caption{Summary of $R$ data and values.}

\begin{tabular}{ccccccccc}
$E_{cm}$ & $N_{had}^{obs}$ & $N_{bg}$ & $L$  &
$\epsilon_{had}$ & $(1+\delta)$ & $R$  &
Stat.  & Sys.  \\
(GeV)  & & & (nb$^{-1})$ & (\%) & &  & error & error \\
\tableline
2.60 & 5617 & 127 & 292.9 & 54.11 & 1.009 & 2.64 & 0.05 & 0.19 \\
3.20 & 2051 & 100 & 109.3 & 65.71 & 1.447 & 2.21 & 0.07 & 0.13 \\
3.40 & 2149 & 178 & 135.3 & 69.33 & 1.173 & 2.38 & 0.07 & 0.16 \\
3.55 & 2672 & 216 & 200.2 & 70.66 & 1.125 & 2.23 & 0.06 & 0.16 \\
4.60 & 1497 & 282 &  87.7 & 81.75 & 1.079 & 3.58 & 0.20 & 0.29 \\
5.00 & 1648 & 463 & 102.3 & 83.94 & 1.068 & 3.47 & 0.32 & 0.29 \\
\end{tabular}
\label{tab:rvalue}
\end{table}

%\vspace{-0.1in}
\begin{table}[!h]
\caption{Contributions to systematic errors: hadronic selection,
$f$ factor, luminosity determination, $\tau$-pair background,
background from Bhabha events, hadronic efficiency determination,
trigger efficiency, and radiative corrections.  All errors are in
percentages (\%).}
\begin{tabular}{ccccccccc}
$E_{cm}$ & Had. & $f$ & $L$ & $\tau$-pair & Bhabhas
& Had. & Trig. & Rad. \\
(GeV) & sel. & factor &  &  &  & eff. & & corr. \\
\tableline
2.60 & 5.1 & 0.06 & 2.12 & 0.00 & 0.04 & 4.10 & 0.50 & 2.6 \\
3.20 & 3.8 & 0.15 & 2.83 & 0.00 & 0.04 & 1.90 & 0.50 & 2.2 \\
3.40 & 4.6 & 0.27 & 2.83 & 0.00 & 0.04 & 2.90 & 0.50 & 3.0 \\
3.55 & 5.5 & 0.27 & 2.32 & 0.00 & 0.04 & 2.30 & 0.50 & 2.4 \\
4.60 & 5.7 & 0.75 & 2.16 & 0.32 & 0.00 & 3.60 & 0.50 & 4.1 \\
5.00 & 6.0 & 1.26 & 2.81 & 0.32 & 0.00 & 3.20 & 0.50 & 3.8 \\
\end{tabular}
\label{tab:rsyst}
\end{table}

%&&&&&&&&&&&&&&&&&&&&&&&&&&&&&&&&&&&&&&&&&&&&&&&&&&&&&&&&&&&&&&&&&&&&&&
%\clearpage
%%%%%%%%%%%%%%%%%%%%%%%%%%%%%%%%%%%%%%%%%%%%%%%%%%%%%%%%%%%%%%%%%%%%%%%

We would like to thank the staff of the BEPC accelerator and IHEP Computing
Center for their efforts.  We also wish to acknowledge useful discussions
with B. Andersson, H. Burkhardt, M. Davier, B. Pietrzyk,
T. Sj\"{o}strand,  M. L. Swartz, J. M. Wu, C. X. Zhang, X. M. Zhang,
and G. D. Zhao.
We especially thank M. Tigner for major contributions
not only to
BES but also to the operation  of the BEPC. 
%%%%%%%%%%%%%%%%%%%%%%%%%%%%%%%%%%%%%%%%%%%%%%%%%%%%%%%%%%%%%%%%%%%%%%%%

\end{document}